# MODELLING OF BOSE–EINSTEIN CONDENSATION IN A WATER TANK

G. Rousseaux[1] and Y. Stepanyants[2,3*)]

[1]Institut Pprime, UPR 3346, CNRS - Université de Poitiers - ISAE ENSMA, 11 Boulevard Marie et Pierre Curie, Téléport 2, BP 30179, 86962 Futuroscope Cedex, France;
[2]University of Southern Queensland, Toowoomba, QLD, 4350, Australia;
[3]Nizhny Novgorod State Technical University n.a. R.E. Alekseev, Nizhny Novgorod, Russia;

## Abstract

It is shown that surface waves propagating against the external current, slowly varying in the horizontal direction in deep water, are governed by the equation which is tantamount to the Gross–Pitaevskii equation modelling the mean-field dynamics of Bose–Einstein condensate. The repulsive or attractive sign of the cubic term in the Gross–Pitaevskii equation is controlled by the choice of the carrier wavelength of the surface waves, while the spatial variation of the current plays the role of the external potential in that equation. The current profile can be easily controlled in the experiments by small variation of the bottom profile, so that the corresponding effective potential in the Gross–Pitaevskii equation can be made in the form of a well or hump. It is shown that the phenomenon of the Bose–Einstein condensation can be effectively emulated in relatively simple laboratory setups for water waves. Generating perturbations with an appropriate carrier wavelength, one can create patterns in the form of trapped waves which correspond to pinned states of Gross–Pitaevskii equation with local potentials. The estimates demonstrate that the parameters of bottom profile, background current, and surface waves are quite accessible to laboratory experiments.

___________________________________________________

*) Corresponding author: Yury.Stepanyants@usq.edu.au.

## I. Introduction

The Bose–Einstein condensation (BEC) has drawn a great deal of attention in course of the last two decades. The condensate effect was experimentally demonstrated in various media, including ultracold atomic and exciton-polariton gases, etc. (see, e.g., Refs. [1–4] and references therein. The use of particular external potentials is a necessary ingredient of these experiments, which are run in sophisticated apparatuses. On the other hand, many dynamical matter-wave regimes characteristic to BEC may be emulated, using simpler equipment, in water-wave tanks. To promote this possibility, in the present work we demonstrate that the basic mean-field BEC model, known as the Gross–Pitaevskii (GP) equation, can be derived for surface water waves in a tank with a spatially varying current. Depending on the wavenumber of the carrier wave, the effective GP equation for the surface waves can be made equivalent to the GP equation with both attractive and repulsive inter-atomic interactions in BEC. External potentials in the GP equation for water waves can be easily emulated in the water tank with an uneven bottom, featuring wells or humps. Particular exact solutions of the effective GP equation are reported here, and estimates for their realization in the water tank are given.

## 2. Derivation of the effective Gross–Pitaevskii equations for waves on the surface of moving water

Following Ref. [5], we consider water-wave propagation on top of a smoothly varying current along the $x$-axis, with flow velocity $U(x) = U_0 + U_1(x)$ including a constant mean value $U_0$ and a small variable component $U_1(x)$, with $\max[|U_1(x)|]/U_0 \ll 1$ (as shown below, the latter term may be induced by a bottom profile of the tank). For a counter-current propagating sinusoudal wave of a small but finite amplitude $A$, with frequency $\omega$ and wavenumber $k$, the dispersion relation for deep water in the laboratory reference frame is [5]:

$$\omega = -U(x)k + \sqrt{gk(1+Tk^2)}\left(1 + \frac{A^2k^2}{2}\right), \tag{2.1}$$

where $g$ is the gravity acceleration, $T = \sigma/\rho g$, $\sigma$ is the surface tension, $\rho$ the water density, and only the term $\sim \varepsilon^2$ with respect to the wave steepnes, $\varepsilon = Ak$, is retained in the respective expression for the Stokes' correction to the wave frequency (see, e.g., Refs. [6, 7]). It is assumed that spatial scale $L$ of the variation of the external current is much greater than the wavelength $\lambda = 2\pi/k$, which makes it meaningful to consider the $x$-dependent frequency in Eq. (2.1). Figure 1 schematically illustrates the respective configuration of the flow and counter-current propagating wave packet.

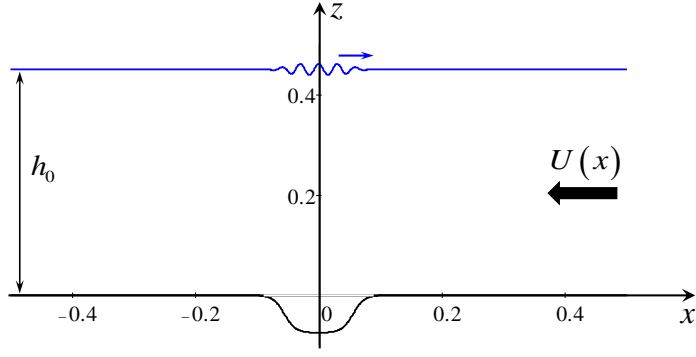

Fig. 1. A sketch of the surface wave train propagating against the current in a water tank with a bottom well.

We consider a weakly modulated wavetrain with the central wave number $k_0$ and frequency $\omega_0 = -U_0 k_0 + \sqrt{g k_0 (1 + T k_0^2)}$. Dispersion relation (2.1) can be expanded around the point ($\omega_0$, $k_0$, $A = 0$) into the Taylor series up to the terms of the order of $\varepsilon^2$ (see Refs. [8, 9, 5]):

$$\omega - \omega_0 = \frac{\partial \omega}{\partial k}(k - k_0) - U_1 k_0 - \frac{1}{2}\frac{\partial^2 \omega}{\partial k^2}(k - k_0)^2 + \frac{\partial \omega}{\partial A^2}(A k_0)^2, \qquad (2.2)$$

where

$$\frac{\partial \omega}{\partial k} = c_g \frac{1 + 3 T k_0^2}{\sqrt{1 + T k_0^2}} - U_0, \quad \frac{\partial^2 \omega}{\partial k^2} = \frac{c_g}{2 k_0} \frac{\left(1 - 3 T k_0^2\right)^2 - 12 T^2 k_0^4}{\left(1 + T k_0^2\right)^{3/2}},$$

$$\frac{\partial \omega}{\partial A^2} = \frac{c_g k_0}{8} \frac{8 + T k_0^2 + 2 T^2 k_0^4}{\left(1 + T k_0^2\right)^{3/2}\left(1 - 2 T k_0^2\right)}(A k_0)^2,$$

where $c_g = (1/2)\sqrt{g_0 k_0}$ is the group velocity of a purely gravity wave with the wavenumber $k_0$ on still water, and

$$\omega - \omega_0 \sim k - k_0 \sim \varepsilon, \quad \max[|U_1(x)|]/U_0 \sim \varepsilon^2. \qquad (2.3)$$

(recall that $\varepsilon \ll 1$ is the small wave steepness).

The evolution equation in the $(x, t)$-space corresponding to dispersion relation (2.2) can be easily restored by replacing $\omega - \omega_0 \to i\,\partial/\partial t$ and $k - k_0 \to -i\,\partial/\partial x$ [10, 11, 5]. Thus we obtain the equation for a slowly varying in space and time complex amplitude of a wave train:

$$i\left(\frac{\partial A}{\partial t} + V_g \frac{\partial A}{\partial x}\right) = U_1(x) k_0 A - \alpha |A|^2 A - \beta \frac{\partial^2 A}{\partial x^2}, \qquad (2.4)$$

where the group velocity of gravity-capillary waves propagating on top of a current is:

$$V_g(k_0, U_0) = c_g \frac{1 + 3 T k_0^2}{\sqrt{1 + T k_0^2}} - U_0, \qquad (2.5)$$

and the coefficients in Eq. (2.4) are

$$\alpha = \frac{c_g k_0^3}{8} \frac{8 + Tk_0^2 + 2T^2 k_0^4}{\left(1 + Tk_0^2\right)^{3/2} \left(1 - 2Tk_0^2\right)}, \tag{2.6a}$$

$$\beta = \frac{1}{2} \frac{d^2 \omega}{dk^2}\bigg|_{k=k_0} = \frac{c_g}{4k_0} \frac{\left(1 - 3Tk_0^2\right)^2 - 12T^2 k_0^4}{\left(1 + Tk_0^2\right)^{3/2}}. \tag{2.6b}$$

A similar equation was derived in Ref. [12] for purely gravity waves (the rigorous derivation of such equation with the help of the asymptotic expansion method can be found in Rev. [5]).

If we choose for the counter-current propagating wave the speed of the underlying current such that $V_g = 0$, i.e. $U_0 = c_g (1 + 3Tk_0^2)/(1 + Tk_0^2)^{1/2}$, then we obtain the standard Gross–Pitaevskii equation [1]:

$$i\frac{\partial A}{\partial t} + \alpha |A|^2 A + \beta \frac{\partial^2 A}{\partial x^2} - U_1(x) k_0 A = 0, \tag{2.7}$$

where $U_1(x)$ plays a role of the external potential, which is shaped as a well, with $U_1(x) < 0$, and as a hump, with $U_1(x) > 0$.

Equation (2.7) without the external potential reduces to the integrable nonlinear Schrödinger (NLS) equation [7–10]. Depending on coefficients $\alpha$ and $\beta$, cnoidal-wave periodic solutions of the NLS equation can be stable or unstable against self-modulating perturbations. According to the Lighthill criterion [7, 9], the stability occurs at $\alpha\beta < 0$, and the instability takes place at $\alpha\beta > 0$. The analysis of coefficients $\alpha$ and $\beta$ shows [8, 9] that, in the case of purely gravity waves, both $\alpha$ and $\beta$ are positive, hence the sinusoidal wavetrains are unstable, when $Tk_0^2 < 2/\sqrt{3} - 1 \approx 0.155$. In the relatively narrow range, $2/\sqrt{3} - 1 < Tk_0^2 < 1/2$, the signs are $\alpha > 0$ and $\beta < 0$, hence the waves are modulationally stable.

Finally, when $Tk_0^2 > 1/2$, both $\alpha$ and $\beta$ are negative, hence the cnoidal waves are again modulationally unstable. The critical wavenumbers, at which coefficients $\beta$ and $\alpha$ change their signs for clean water at temperature 25°C are, respectively, $k_1 = 1.452$ cm$^{-1}$ ($\lambda_1 = 4.33$ cm) and $k_2 = 2.61$ cm$^{-1}$ ($\lambda_1 = 2.41$ cm). Note that the group velocity in still water, $V_g(k, 0)$, attains a minimum at the former critical point $k_1$ [7]. At both critical points $k_{1,2}$, Eq. (2.7) should be replaced with a more complex equation [5] (we do not consider such degenerate cases in detail here. Figure 2 shows the intervals of wavenumbers where the modulational stability and instability occur.

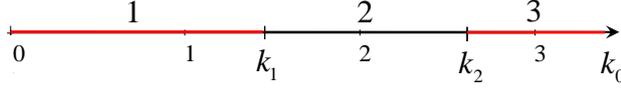

Fig. 2. Intervals of instability (1 and 3) and stability (2) for sinusoidal surface waves.

In the next sections we demonstrate that, using various shapes of the bottom well, one can produce different corrections to the main flow $U_1(x)$ in a water tank, via the conservation of the flow rate through the tank's cross-section, $U_1(x)h_1(x)$ = const. For some shapes of the bottom well, exact solutions of the effective GP equation can be constructed both for the repulsive and attractive signs of the nonlinear term. It is easy to make bottom wells of various shapes in the tank, and trapped surface waves on top of the corresponding current can be readily observed in the experiment.

## 3. An example of exact solution of the effective Gross–Pitaevskii equation in the modulationally stable case

In this section we demonstrate that one of basic exact solutions of the GP equation can be realized in the laboratory experiment with modulationally stable surface waves belonging to the interval 2 shown in Fig. 2. To this end, we first assume the presence of a constant water flow $U_0$ = 0.185 m/s in the tank of constant depth $h_0$ = 0.45 m. Figure 3 displays the dispersion relation (2.1) for surface waves of infinitesimal amplitude (with $A = 0$) and constant current speed $U_0$.

Further, we assume the presence of a shallow well in the central part of the tank's bottom, which modifies the total depth so that

$$h(x) = h_0 \left\{ 1 + F \left[ \tanh\left(\frac{x}{\Delta} + \phi\right) - \tanh\left(\frac{x}{\Delta} - \phi\right) \right] \right\}, \quad (3.1)$$

where $\phi = \frac{1}{4}\ln\frac{1+\nu}{1-\nu}$, $\Delta = \sqrt{\frac{-6\beta}{\nu F k_0 U_0}}$, $F > 0$ and $\nu$ being free parameters ($0 < \nu < 1$), which control the depth of the cavity and its shape. The largest variation of the depth, corresponding to Eq. (3.1), is

$$\delta h = h(0) - h_0 = 2h_0 F \frac{1 - \sqrt{1-\nu^2}}{\nu}. \quad (3.2)$$

The front and rear slopes of the well, $\Delta$, depend monotonically on parameter $\nu$, decreasing from infinity to $\Delta_{\min} = \sqrt{-6\beta/(Fk_0U_0)}$, when $\nu$ varies from 0 to 1. The characteristic width of the well, $L$, i.e., the distance between its frontal and rear segments at the half-maximum level, $\delta h/2$, is

$$L = \sqrt{\frac{-6\beta}{vFk_0U_0}} \ln\left(\frac{2\sqrt{1-v^2}+1+\sqrt{4-3v^2+4\sqrt{1-v^2}}}{\sqrt{1-v^2}}\right). \tag{3.3}$$

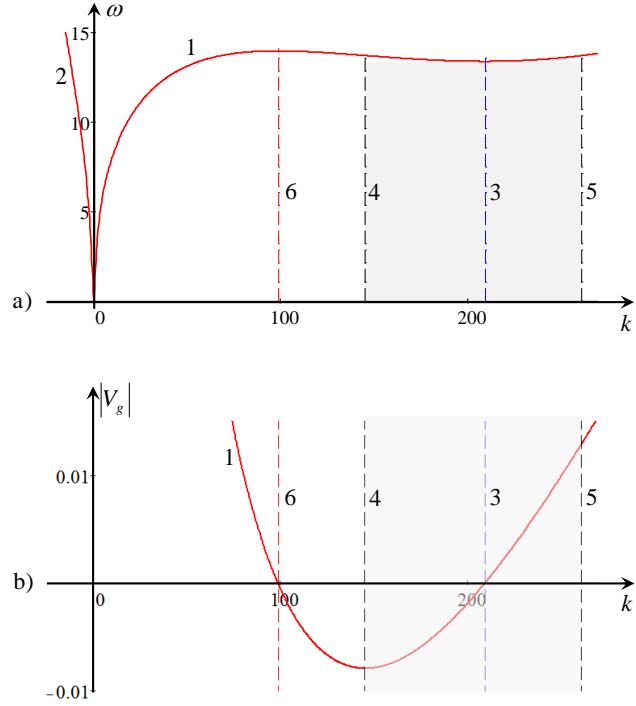

Fig. 3. The dispersion relation for surface waves with an infinitesimal amplitude. Frame a) shows $\omega(k)$ as per Eq. (2.1) with $A = 0$; frame (b) shows the group velocity in the presence of the underlying current. Lines 1 and 2 pertain, respectively, to counter- and co-current propagating waves (the group velocity for the latter branch is not shown). Vertical dashed lines 3 and 6 correspond to the carrier waves with $k_0 = 2.1$ cm$^{-1}$ and $k_0 = 0.985$ cm$^{-1}$, respectively; dashed lines 4 and 5 show the boundaries of the shaded domain where sinusoidal waves are modulationally stable. Line 4 corresponds to $k_1$, and line 5 − to $k_2$, as per Fig. 2.

At $v \to 0$, the well takes the shape of an inverted bell, whose width increases as $1/\sqrt{v}$: $L \approx \sqrt{\frac{-6\beta}{vFk_0U_0}} \ln(3+\sqrt{8})$. In another limit, $v \to 1$, the well becomes very wide too, with the width increasing as $L \approx \sqrt{\frac{-3\beta}{2Fk_0U_0}} \ln\frac{2}{1-v}$. The minimum width, $L_{\min} \approx 2.2\sqrt{\frac{-6\beta}{Fk_0U_0}}$, is attained at $v \approx 0.821$. All such shapes can be readily designed in the experimental setup.

Due to the conservation of the mass flux through any cross-section, the variation of the depth causes the variation of the speed, therefore, above the bottom well, the current varies as follows:

$$U(x) = \frac{U_0}{1 + F\left[\tanh\left(\frac{x}{\Delta} + \phi\right) - \tanh\left(\frac{x}{\Delta} - \phi\right)\right]} \approx U_0\left\{1 - F\left[\tanh\left(\frac{x}{\Delta} + \phi\right) - \tanh\left(\frac{x}{\Delta} - \phi\right)\right]\right\}, \quad (3.4)$$

where $F \ll 1$ is assumed, hence the effective potential in Eq. (2.7) is:

$$U_1(x) = -U_0 F\left[\tanh\left(\frac{x}{\Delta} + \phi\right) - \tanh\left(\frac{x}{\Delta} - \phi\right)\right]. \quad (3.5)$$

To realize the dynamical regime corresponding to the GP equation with the self-attraction, we chose a surface mode with wavelength $\lambda_0 = 3$ cm ($k_0 = 2\pi/\lambda_0 = 2.1$ cm$^{-1}$) and amplitude $A \approx 1$ mm, hence the corresponding wave steepness, $\varepsilon = Ak_0 = 0.21$, may be considered as a small parameter. For such a wave, even in a relatively shallow section of the tank we have $k_0 h_0 = 135 \gg 1$, which means that the deep-water case realizes. The absolute value of the group velocity of such a wave in the absence of the underlying current is $V_g(k_0, 0) = 18.5$ cm/s, whereas the minimal group velocity for given parameters is $V_{\min} = -0.79$ cm/s, see Fig. 3b. An obviously interesting possibility is to observe a "standing water soliton", i.e., to bring the wavetrain, travelling counter-current, to a halt in the laboratory frame. To this end, we set $U_0 = c_g\left(1 + 3Tk_0^2\right)/\sqrt{1 + Tk_0^2}$; according to Eqs. (2.5), and (2.6) this determines the nonlinearity and dispersion coefficients in GP equation (2.7): $\alpha = 3.348 \cdot 10^2$ cm$^{-2}\cdot$s$^{-1}$, $\beta = -1.806 \cdot 10^2$ cm$^2\cdot$s$^{-1}$.

As has been shown in Ref. [13], the GP equation with the potential given by Eq. (3.5) admits the exact solution in the form of

$$A(x,t) = \exp(-i\Omega t)\sqrt{\frac{\nu F k_0 U_0}{3\alpha}}\left[\tanh\left(\frac{x}{\Delta} + \phi\right) - \tanh\left(\frac{x}{\Delta} - \phi\right)\right], \quad (3.6)$$

where $\Omega = 2\nu F k_0 U_0/3$ is a nonlinear correction to the wave frequency $\omega_0$, and the amplitude of the localized state is

$$A_{\max} = 2\sqrt{\frac{F k_0 U_0}{3\alpha\nu}}\left(1 - \sqrt{1 - \nu^2}\right). \quad (3.7)$$

The total norm of this solution (which gives a scaled number of atoms in the application to BEC) is:

$$N = \int_{-\infty}^{+\infty} |A(x)|^2\, dx = \frac{4\nu F k_0 U_0 \Delta}{3\alpha}\left[2\phi\coth(2\phi) - 1\right]. \quad (3.8)$$

The normalized squared absolute value of solution (3.6), corresponding to the local density of atoms in BEC, along with the normalized potential, $U_1(x)/(U_0 F)$, are shown in Fig. 4 for several values of free parameter $\nu$. As demonstrated in Ref. [13], this exact solution is actually the ground state of the GP equation with the repulsive nonlinearity and potential well (3.5), hence this solution is definitely stable.

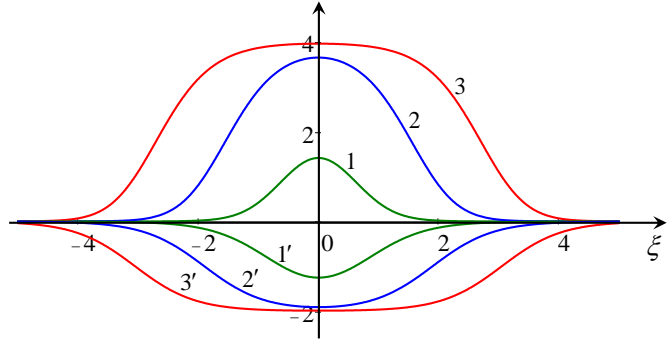

Fig. 4. Normalized solution (3.6) in terms of $3\alpha|A(x)|^2/(FU_0k_0)$ (lines labelled 1, 2, 3), and the corresponding normalized potentials $U_1(x)/(U_0F)$ (lines labelled 1′, 2′, 3′) as functions of dimensionless coordinate $\xi = x/\Delta$. Lines 1 and 1′ pertain to $\nu = 0.9$; lines 2 and 2′ to $\nu = 0.999$; lines 3 and 3′ to $\nu = 0.99999$.

Thus, a surface gravity-capillary wave with carrier wavelength $\lambda_0 = 3$ cm can be trapped in the water flow over the bottom well. If one takes, for example, the value of the free parameter $F = 0.15$, the largest depth of the well at $\nu = 0.999$ is $\delta h \equiv h_{max} - h_0 = 12.9$ cm, so that $\delta h/h_0 = 0.29$. The width of the well is $L \approx 53$ cm, according to Eq. (3.2), the envelope of the trapped wavetrain having the same width, see Fig. 4. The amplitude of the wave is $A_{max} = 1.5$ mm, hence $(k_0A_{max})^2 = 0.3$, whereas $(|U_1(x)/U_0|)_{max} = 0.29$, which agrees with the assumption of the smallnesses of the wave steepness and modulation of the basic current, $(|U_1(x)/U_0|)_{max}$ and $\varepsilon^2$ being of the same order of magnitude, in agreement with Eq. (2.3).

It has been shown in Ref. [13] that there are many other exact stable solutions of the GP equation with the corresponding potentials, which can be easily realized in the water tank.

## 4. An example of exact solution of the effective Gross–Pitaevskii equation in the modulationally unstable case

We now consider the situation with a small bell-shaped well in the central part of the tank's bottom which modifies the depth as [cf. Eq. (3.1), which represented another well's profile]

$$h(x) = h_0\left[1 + \frac{F}{1 + B\cosh(x/\Delta)}\right], \quad (4.1)$$

where $\Delta$ and $B > 1$ are free parameters, which control the depth and width of the cavity, and $F$ is a function of $\Delta$, to be specified below. The maximal variation of well's depth is:

$$\delta h \equiv h(x) - h_0 = \frac{h_0 F}{1+B}. \tag{4.2}$$

where it is assumed that second term in the square brackets is small in comparison with 1.

To consider the effective GP equation with the self-attraction, i.e., modulational instability of the surface wave, we choose it with wavelength $\lambda_0 = 6.4$ cm ($k_0 = 2\pi/\lambda_0 = 0.985$ cm$^{-1}$) and amplitude $A \approx 3.24$ mm, so that the wave steepness $\varepsilon = Ak_0 \approx 0.32$ may again be treated as a small parameter. For such a wave, even in a relatively shallow section of the tank we have $k_0 h_0 = 44.33 \gg 1$, which means that the deep-water approximation remains valid. The group velocity of such a wave in the absence of the underlying current is the same as in the previous example, viz., $V_g(k_0, 0) = 18.5$ cm/s, see Fig. 3(b). To bring a counter-current traveling wavetrain to a halt in the laboratory frame (as done above, to produce a "standing water hump") we again set $U_0 = V_g(k_0, 0)$, pursuant to Eq. (2.5). Then Eq. (2.6) produces the nonlinearity and dispersion coefficients of GP equation (2.7): $\alpha = 18.6$, cm$^{-2}\cdot$s$^{-1}$, $\beta = 2.385\cdot 10^2$, cm$^2\cdot$s$^{-1}$, and the effective potential,

$$U_1(x) = \frac{-U_0 F}{1 + B\cosh(x/\Delta)}. \tag{4.4}$$

It is easy to check that the GP equation with this potential and self-attractive cubic term has an exact localized solution (a soliton pinned to the potential well) in the form of

$$A(x,t) = \frac{R\exp(-i\Omega t)}{1 + B\cosh(x/\Delta)}, \tag{4.5}$$

where $\Omega = \beta/\Delta^2$ is a nonlinear correction to the wave frequency $\omega_0$, and $R$ and $F$ are expressed in terms of free parameters $B$ and $\Delta$:

$$R = \frac{1}{\Delta}\sqrt{\frac{2\beta}{\alpha}(B^2 - 1)}, \quad F = \frac{-3\beta}{U_0 k_0 \Delta^2}. \tag{4.6}$$

The amplitude of this pinned soliton is

$$A_{\max} = \frac{R}{1+B} = \frac{1}{\Delta}\sqrt{\frac{2\beta}{\alpha}\frac{B-1}{B+1}}. \tag{4.7}$$

The stability of this solution was verified in Ref. [13]. The profile of its squared absolute value, corresponding to the local density of atoms in BEC, along with the respective normalized potential, $U_1(x)/(U_0 b)$, are shown in Fig. 5 for dimensionless parameter $B = 2.5$ and $\Delta = 25$ cm.

Thus, we see that the surface gravity wave with the carrier wavelength $\lambda_0 = 6.4$ cm can be trapped in the water flow over the bottom well considered here. For the chosen free parameters $B$ and $\Delta$ we find that the maximal depth of the well is $\delta h \equiv h_{\max} - h_0 = 0.8$ cm ($\delta h/h_0 \approx 0.018$). The number of periods of the carrier wave within the envelope of the localized trapped mode is $2\Delta/\lambda_0$ ≈ 8. The same width has the envelope of the trapped wave train. The amplitude of the mode is $A_{max}$ ≈ 1.3 mm, hence the corresponding wave steepness is $\varepsilon \equiv k_0 A_{max}$ ≈ 0.13. The largest variation of the mean flow, induced by the bottom well, is $|U_1(x)/U_0|_{max}$ ≈ 0.018. This agrees well with our underlying assumptions about the smallnesses of the wave steepness, and $|U_1(x)/U_0|_{max}$ ~ $\varepsilon^2$, see Eq. (2.3).

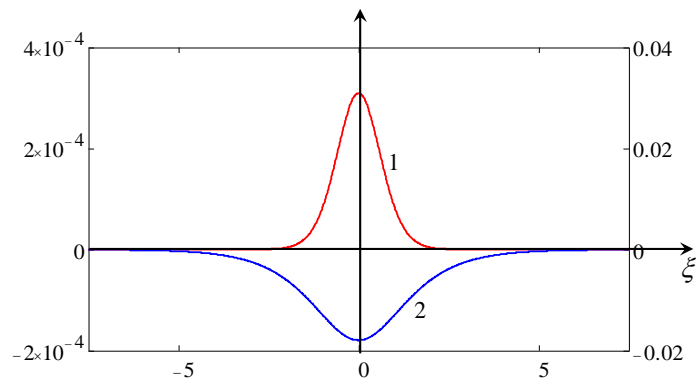

Fig. 5. The squared absolute value of solution (4.5) (line 1) and the corresponding normalized potential $U_1(x)/(U_0)$ (line 2). as functions of dimensionless coordinate $\xi = x/\Delta$. The left and right scales pertains, respectively, to lines 1 and 2.

In the case of purely gravity waves, when the GP equation has the attractive nonlinearity, many other exact solutions with the corresponding potentials are available [13]; they can be relatively easy realized in the water tank, using the surface waves with $\lambda$ > 4.3 cm.

## 5. Conclusion

In this paper we have shown that surface waves propagating against the external current, slowly varying in the horizontal direction in deep water, are governed by the equation which is tantamount to the GP (Gross–Pitaevskii) equation modeling the mean-field BEC dynamics. The repulsive or attractive sign of the cubic term is controlled by the choice of the carrier wavelength of the surface waves, while the spatial variation of the current plays the role of the external potential in the GP equation. The current profile can be easily controlled in the experiments by small variation of the bottom profile, so that the corresponding effective potential in the GP equation can be made in the form of a well or hump.

For some particular bottom profiles the effective GP equation admits exact solutions, which can be experimentally implemented in the water tank with the background current. Thus, the phenomenon of the Bose–Einstein condensation can be effectively emulated in relatively simple laboratory setups for water waves. Generating perturbations with an appropriate carrier wavelength, one can create patterns in the form of trapped waves which correspond to pinned states of the GP equation with local potentials. Our estimates presented in the paper demonstrate that the parameters of bottom profile, background current, and surface waves are quite accessible to laboratory experiments.

Lastly, we note that, formally speaking, in addition to what is elaborated above, the GP equation can be also implemented in the domain 3 in Fig. 4 for purely capillary waves with $\lambda < 2.4$ cm. However, for such short waves water viscosity becomes important, which would complicate the comparison between the theoretical results predicted by the GP equation and observations.

**Acknowledgements.** The authors are thankful to B. Malomed for the comments and helpful advices. This work was initiated when one of the authors (Y.S.) was an invited Visiting Professor at the Institut Pprime, Université de Poitiers in August–October, 2016 thankful to the grant from Region Poitou–Charentes. G.R. acknowledges funding from the ANR grant HARALAB No ANR-15-CE30-0017-04.

**References**

1. Pitaevskii, L.P., and Stringari, S. (2003). *Bose–Einstein Condensation*, Clarendon, Oxford.
2. Kasprzak, J., Richard, M., Kundermann S., Baas, A., Jeambrun, P., Keeling, J.M.J., Marchetti, F.M., Szymańska, M.H., André, R., Staehli, J.L., Savona, V., Littlewood, P.B., Deveaud, B., and Le Si Dang. (2006). Bose–Einstein condensation of exciton polaritons. Nature, v. 443, 409–414.
3. Byrnes, T., Kim, N. Y., and Yamamoto, Y. (2014). Exciton–polariton condensates. Nature Phys., v. 10, 803–813.
4. Pethick, C.J., and Smith, H. (2008). *Bose–Einstein Condensation in Dilute Gases*. Cambridge University Press, 2nd Edition.
5. Stocker, J.R., and Peregrine, D.H. (1999). The current-modified nonlinear Schrödinger equation. J. Fluid Mech., v. 399, 335–353.
6. Lamb, H. (1932). *Hydrodynamics*, 6th edn. Cambridge University Press.
7. Whitham, G.B. (1974). *Linear and nonlinear waves*, Wiley Interscience Publ., John Wiley and Sons, N. Y.

8. Djordjevic, V.D., and Redekopp, L.G. (1977). On two-dimensional packets of capillary-gravity waves. J. Fluid Mech., v. 79, pt. 4, 703–714.

9. Ablowitz, M.J., and Segur, H. (1981). *Solitons and the Inverse Scattering Transform*, SIAM, Philadelphia.

10. Yuen, H.C., and Lake, B.M. (1982). Nonlinear dynamics of deep-water gravity waves. Adv. Appl. Mech., v. 22, 67–229.

11. Korpel. A., and Banerjee, P.P. (1984). A heuristic guide to nonlinear dispersive wave equations and soliton-type solutions. Proc. IEEE, v. 72, n. 9, 1109–1130.

12. Bakhanov, V.V., Kemarskaya, O.N., Pozdnyakova, V.I., Okomel'kova, I.A., and Shereshevsky, I.A. (1996). Evolution of surface waves of finite amplitude in field of inhomogeneous current. Proc. Intl. Geoscience & Remote Sensing Sympos. v. 1, 609–611.

13. Malomed, B.A., and Stepanyants, Y.A. (2010). The inverse problem for the Gross−Pitaevskii equation. Chaos, v. 20, 013130, 14 p.